\documentstyle[11pt]{article}
\input{epsfig.sty}

\newcommand{\cA}{{\cal A}}
\newcommand{\be}{\begin{equation}}
\newcommand{\ee}{\end{equation}}
\newcommand{\ba}{\begin{array}}
\newcommand{\ea}{\end{array}}

\begin{document}

\begin{titlepage}

\pagestyle{plain}

\title{ 
{\bf Isospin and density dependences of nuclear matter 
symmetry energy coefficients 
 } }
\author{  {\bf F\'abio L. Braghin} \thanks{e-mail: 
braghin@if.usp.br}  \\
{\normalsize 
 Nuclear Theory and Elementary Particle
 Phenomenology Group,}\\
{\normalsize 
Instituto de F\'\i sica, Universidade de 
S\~ao Paulo, C.P. 66.318; CEP 05315-970;  S\~ao Paulo - SP, Brazil. }\\
}

\maketitle
\begin{abstract} 
Symmetry energy coefficients of explicitly isospin asymmetric nuclear matter
at variable  densities (from .5$\rho_0$ up to 2 $\rho_0$)
are studied as generalized screening functions.
An extended stability condition for asymmetric nuclear matter
is proposed. 
We find the possibility of obtaining stable asymmetric nuclear matter 
even in some cases for which 
the symmetric nuclear matter limit is  unstable.
Skyrme-type forces are extensively used in 
 analytical expressions of the symmetry energy coefficients
derived as generalized screening functions 
in the four channels of the particle hole interaction
 producing alternative behaviors at different
$\rho$ and $b$ (respectively the density and the asymmetry coefficient).
The spin and spin-isospin coefficients,
with corrections to the usual Landau Migdal parameters, 
indicate the possibility of occurring instabilities with common features
depending on the nuclear density and n-p asymmetry.
Possible relevance for high energy heavy ions collisions and 
astrophysical objects is discussed.
\end{abstract}

PACS numbers: 21.30.-x, 21.65.+f, 26.50.+x, 26.60.+c

Key-words: Symmetry energy, nuclear density, isospin asymmetry,
screening functions, isospin, spin, incompressibilities.

IF- USP - 2001, 2003

\end{titlepage}

\section{Introduction}

The symmetry energy coefficients and their dependences on the
density and isospin symmetry are of relevance, for example,
for the description of macroscopic nuclear properties
as well as for 
proto-neutron and neutron stars.
In the case of the neutron-proton (n-p) 
symmetry energy coefficient (s.e.c) of nuclear matter 
($a_{\tau}$) it represents the 
tendency of nuclear forces to have
greater binding energies (E/A) for symmetric systems - equal number of
protons and neutrons.
It contributes as a  coefficient for the squared 
neutron-proton asymmetry in usual macroscopic mass formula
in the parabolic approximation: 
\be \label{1}
E/A = H_0(A,Z) + a_{\tau}(N-Z)^2/A^2,
\ee
where $H_0$  does not depend on the asymmetry,
Z, N and A are the proton, neutron and mass numbers respectively. 
Higher orders effects of the asymmetry (proportional to $(N-Z)^n$
for $n \neq 2$ \cite{JACO}) 
are usually expected to be less relevant
for the equation of state (EOS) of nuclear
matter  based on such parameterizations \cite{LKLB,MONI}.
$a_{\tau}$ is also the parameter
which measures the response of the system to a perturbation which
tends to separate protons from neutrons. 
It is
given by the static polarizability of the system 
(the inverse isospin screening function)
which also should depend on the asymmetry of the medium. 
In the scalar channel one can define the (dipolar) incompressibility 
which is related to the nuclear matter incompressibility \cite{ISOSYMEN}.
These two coefficients (isovector and scalar) and their 
dependences on the neutron-proton asymmetry were initially investigated 
in a previous work \cite{ISOSYMEN}.
Other symmetry coefficients may also be defined in 
nuclear matter, for instance, the
 spin $a_{\sigma}$ and  spin-isovector
$a_{\sigma \tau}$ ones. 
The former would
correspond to the difference in the binding energy due
 the inclusion of a polarized nucleon in the medium
whereas the second involves  the distinction between 
neutrons and protons as well.
A calculation for the dynamical and static  polarizabilities -
proportional to the inverse of such
symmetry coefficients in asymmetric
matter - was done using Skyrme effective forces 
in \cite{ISOSYMEN,FLB99,BJP2003}.
The spin channel is relevant for the study of the neutrino interaction
with matter because it couples with axial vector current together with 
the scalar channel \cite{SAWYER,ESPANHOIS,REDDY}. 
A suppression of the spin susceptibility 
(in this work we will be dealing rather with its inverse,
 the spin symmetry coefficient - $A_{1,0}$, as 
shown in sections 2 and 3,
corresponding also to an increase of the Landau parameter $G_0$)
lead to the suppression of Gamow Teller transitions which are of interest 
for the supernovae mechanism \cite{REDDY}. 
Calculations with Skyrme interactions usually result in smaller $G_0$
(corresponding to smaller spin symmetry coefficients) than microscopic
calculations leading to instabilities associated to 
ferromagnetic polarized states \cite{KUTSCHERA,REDDY}.
In the present work we show this  conclusion about Skyrme interactions
is not necessarily correct.
The spin-isospin channel has been associated with instabilities 
which would lead to pion condensation. 
In the language of Migdal-Landau parameters one would have 
the parameter $G_0' < -1$  for the formation of such condensates. 

In the present work we show that these coefficients may
provide us with a way of checking the stability of
asymmetric nuclear matter with respect to the explicit isospin asymmetry.
Another aim
is to articulate the idea that different 
 Skyrme-type forces may be appropriate to 
the description of diverse phenomena at variable ranges
of the nuclear density and asymmetry by extending
the previous studies of s.e.c.
Some Skyrme interactions will be used to 
assess the possible behavior of these functions.
The parameters of one of the forces we use (SLyb) were fitted from  results 
of neutron matter properties obtained from microscopic calculations 
in \cite{CHABANAT}. Other forces (SkSC4, SkSC6 and SkSC10), which
have slightly different density dependencies, had their parameters
fixed by adjusting a large amount of nuclear masses 
\cite{DUTOABO,ONSIPP}.

This paper reviews and extends the works presented in \cite{ISOSYMEN,BJP2003} 
and 
it is organized as follows.
In the next section we remind an argument
previously developed
to investigate generalized static polarizabities  
which allow for the study of the nuclear matter stability.
In section 3 the static polarizabilities in the four channels of the 
particle-hole interaction with several Skyrme forces is studied in asymmetric
nuclear matter at variable densities
as derived from the linear response method.
Next a new stability condition 
with respect to neutron-proton fluctuations asymmetry 
is proposed.
In the last section possible consequences are pointed out and
the  results are summarized.

\section{Generalized Symmetry Energy Coefficients}

We review a qualitative argument  from  
\cite{BVA,ISOSYMEN} for exploring them.
We consider a small amplitude ($\epsilon$) 
 external perturbation which acts, through the third
Pauli isospin matrix  $\tau_3$, in nuclear matter 
 separating nucleons with isospin up and down 
\footnote{
This argument is valid for the four channels. 
It is enough to consider other external 
perturbations: $\sigma_3$, $\sigma_3 \tau_3$ and {\bf 1} for 
the spin,  spin-isospin and 
scalar channels respectively.}. 
This originates fluctuations 
$\delta \rho = \delta \rho_n - \delta \rho_p \equiv \beta$ 
of the nucleon densities. 
The energy of the system can be written as:
\be \label{1as}
\displaystyle{ H = H_0 + A_{0,1}\frac{(\rho_n-\rho_p)^2}{\rho} +
\epsilon \beta ,}
\ee
where $\cA_{s,t}$  is the isovector
symmetry coefficient ($s=0,t=1$ - spin, isospin) usually 
denoted by $a_{\tau}, J$ or $\beta$.
For the other channels $(s,t)$ one defines
different symmetry coefficients.
We can re-write
the n-p density difference in terms of $\beta$ as: 
$ \rho_n - \rho_p =  (\rho_{n,(0)}-\rho_{p,(0)} + \beta) $, where 
the subscript $_{(0)}$ indicates static densities, without the 
external source. Although one is dealing with an infinite
system, the arguments are to be valid for finite nuclei.

The polarizability is defined as the ratio of the density 
fluctuation to the amplitude of the external perturbation
and it can be written as \cite{ISOSYMEN,BVA}:
\be \label{1bs} \ba{ll}
\displaystyle{ \Pi^{s,t} \equiv \frac{\beta}{\epsilon }
= - \frac{\rho}{2 \cA_{s,t} (b,\beta) },
}
\ea
\ee
where we consider the general  channel (s,t) 
as done in \cite{FLB99}. 
Note that $\cA_{s,t}$ is a function of $b$ and $\beta$ and these 
parameters may be related, as argued below. 
The occurrence of these functional dependences of 
$\cA_{s,t}$ can be found just by 
the stability condition with respect to the fluctuation 
from expression (\ref{1as}):
\be \label{minimH} \ba{ll}
\displaystyle{ \frac{d H}{d \beta} = 0.}
\ea
\ee

\subsection{ Isospin dependence of $\cA_{s,t}$}

The neutron proton asymmetry used in the present 
work is defined by the neutron and proton densities
$\rho_n, \rho_p$ as:
\be \ba{ll}
\displaystyle{ b = \frac{\rho_n}{\rho_p} - 1.}
\ea
\ee
It varies from $b=0$, in symmetric nuclear matter, up to infinity, 
in neutron matter.
The coefficient $b$ is related to a frequently used  
asymmetry coefficient:
\be \label{alphaB} \ba{ll} 
\displaystyle{ \alpha = \frac{(2\rho_{0n}-\rho_0)}{\rho_0},}
\ea
\ee
by the expression: 
\be \label{alphabb} \ba{ll}
\displaystyle{ b = 2\alpha / (1- \alpha ) .}
\ea
\ee
As stated above 
$\cA_{s,t}$ is  a function of the density fluctuation $\beta$.
Although $\beta$ is not the 
explicit n-p asymmetry itself (given by $b$) we will consider that
it depends on it (as it was also argued in \cite{ISOSYMEN,BJP2003}). 
We consider, as shown below, these parameters are related to each other
and therefore we will write ${\cA} = \cA (\beta)$ shortly.
In \cite{FLB99,BJP2003}  different prescriptions were discussed for
$\beta = \beta (b)$ in the calculation of the response function of asymmetric 
nuclear matter. 
We have used (and it was shown to be a reasonable prescription 
for the dynamical response function \cite{FLB99}) 
the  one which leads to the following relation
between  the fluctuation $\beta$ and the 
explicit n-p asymmetry $b$: 
\be \label{relac} \ba{ll}
\displaystyle{ \beta = \delta \rho_n \left( \frac{2+b}{1+b} \right),}
\ea
\ee
Where $\delta \rho_n$ is the neutron density fluctuation. 
In the n-p symmetric limit $\beta = 2 \delta \rho_n$ and in another limit, 
in neutron matter, $\beta = \delta \rho_n$.
The above prescription (expression (\ref{relac})) 
is based on the assumption that
the density fluctuations are proportional to the respective 
density of neutrons and protons, i.e., 
$\delta \rho_n/ \beta = \rho_n / \rho$, being $\rho$ the total 
density.
In spite of being rather appropriated for the isovector channel, 
this kind of assumption can be considered as a starting point 
for the 
other channels (spin, scalar) in asymmetric nuclear matter.
Prescription (\ref{relac})
is therefore model-dependent and different choices for it yield other
 forms for
the (asymmetric) static screening functions.
The  dynamic response functions are less sensitive to this
prescription \cite{FLB99}.

>From the solution of the polarizability (\ref{1bs}) we calculate
the first derivative with relation to $b$:
\be \label{DERIV1} \ba{ll}
\displaystyle{ \frac{d \beta}{d b} = \frac{\epsilon \rho}{2 \cA^2_{0,1} } 
\frac{d \cA^{0,1} }{d b} = - \frac{\beta}{\cA_{0,1}} \frac{d \cA^{0,1}}{d b}. }
\ea
\ee
Another expression can be obtained from the relation between
$b$ and $\beta$ of (\ref{relac}). It yields:
\be \label{relder} \ba{ll}
\displaystyle{ \frac{d \beta}{d b} = - \frac{\beta}{(2+ b)(1+ b)}.}
\ea
\ee
Equating these two last equations we obtain:
\be \label{relder2} \ba{ll}
\displaystyle{  - \cA^{0,1} \frac{\beta}{(2+ b)(1+ b)} = 
- \beta \frac{d \cA^{0,1}}{d b}
,}
\ea
\ee
>From which it is possible to  derive the following relation 
between the isospin s.e.c. and the n-p asymmetry
\cite{ISOSYMEN}:
\be \label{isosb} \ba{ll} 
\displaystyle{ {\cal A}^{0,1} = {\cal A}^{0,1}_{sym} \frac{ 2 + 2 b}{2 + b} 
.}
\ea
\ee
In this expression  $\cA_{sym} = a_{\tau} \simeq 30 MeV$ is the 
s.e.c. of symmetric nuclear matter ($b=0$).
For $b= 2$ (neutron density three times larger than the
proton density) we obtain $\cA = 1.5 \cA_{sym}$. In the
limit of neutron matter 
${\cal A} (b \to \infty) = 2 {\cal A}_{sym}$.
The prescription 
of expression (\ref{relac}) was the relevant information for this calculation. 
Any other relation between $b$ and $\beta$
will induce different asymmetry dependence of the 
symmetry energy coefficient \cite{BJP2003}. 
Another assumption for deriving expression (\ref{isosb}) was that 
$\rho$ is independent of $b$. 
This would be non trivially different if one considers a complete self consistent
calculation with the equation of state of a proto-neutron star, for
example.

\section{ Generalized Screening functions with Skyrme forces }

In this section in in the next one we nearly 
extend the analysis of \cite{ISOSYMEN}.
A nearly exact expression for the dynamical polarizability 
of a non relativistic hot asymmetric nuclear 
matter  at variable densities was derived with Skyrme 
interactions in \cite{FLB99}. 
The approximations were:
 (i) to equate the asymmetry coefficient defined for the
momentum density to the density asymmetry coefficient
(the variation of the corresponding term yields differences 
in the response function of less than $1\% $), 
(ii)
to choose a particular (discussed and justified 
\cite{FLB99}) prescription for the asymmetry density fluctuations
- expression (\ref{relac}).

The generalized static screening functions $A_{s,t}$ (or symmetry 
energy coefficients according to the discussed before)
in asymmetric nuclear matter at finite temperature can be written as
\cite{FLB99}:
\be \label{20} \ba{ll}
\displaystyle{A_{s,t} = 
\frac{\rho}{ N}
\left\{ 1 + 2\overline{V_0^{s,t}} N_c 
+6 V_1^{s,t}M_p^* ( {\rho}_c + {\rho}_d ) + 
12M^*_p V_1^{s,t} \overline{V_0^{s,t}} \left( N_c {\rho}_d
 - {\rho}_c {N}_d \right) + \right.  } \\
\displaystyle{ \left. (V_1^{s,t})^2 \left( 36(M^*_p)^2 {\rho}_c
{\rho}_d - 16M_p^* M_c  N_d \right)  \right\} }.
\ea
\ee
The densities $\rho_v$, $M_v$ and $N_v$  are given respectively by:
 $$ \rho_v = v \rho_n + (1-v) \rho_p, \;\;\;\;
 M_v = v M_n + (1-v) M_p \;\;\;\; N_v = v N_n + (1-v) N_p,$$ 
where $v$ stands for n-p asymmetry coefficients ($c,d$)
defined below (a measure of the fraction of neutron densities). 
The above densities at finite temperature are defined by:
$$ (N_q,\rho_q,M_q) = \frac{2M^*_p}{\pi^2}
\int d f_q(k) (k, k^3, k^5).$$
In these expressions $f_q(k)$ are the
fermion occupation numbers for neutrons ($q=n$) and 
protons ($q=p$) which will be considered only for the
zero temperature limit. 
In this case: $ d f_q(k) = \delta( k - k_F^{q} ) dk_q $, which 
makes the above integrals trivial.
 $\overline{V_0}$ 
and $V_1$ are functions of the
Skyrme forces parameters (see in \cite{FLB99}) and
$M^*_p = m^*_p/(1+a/2)$ is a modified
effective mass for the proton. Besides that, the
four asymmetry coefficients  are:
\be \label{5b} \ba{ll}
\displaystyle{
a = \frac{m^*_{p}}{m^*_{n}} - 1 , \;\;\;\;\;
b = \frac{\rho_{0n}}{\rho_{0p}}  -1, \;\;\;\;\; 
c = \frac{1+b }{2+b }, }\;\;\;\;\;
\displaystyle{ d = \frac{1 }{1+ (1+b)^{\frac{2}{3} } } }.
\ea
\ee


\subsection{Results for Skyrme interactions}

In this section we present results for the s.e.c. 
using Skyrme forces with the expression (\ref{20}) in
all the four channels of the particle-hole interaction.
The scalar, isovector and spin  channels were initially discussed 
in a previous works \cite{ISOSYMEN,BJP2003} and those analysis is extended here.
The present analysis on the spin and spin-isospin channels was not done before.

In Figure 1 we show the generalized isovector symmetry energy
coefficient $A_{0,1}$ 
as a function of the ratio of the nuclear density to the saturation density 
($\rho /\rho_0$)
for Skyrme interactions 
SLyb \cite{CHABANAT}, SkSC4, SkSC6 and SkSC10 \cite{DUTOABO,ONSIPP}. 
Since Skyrme forces are not necessarily expected 
to describe Physics at high densities we 
investigate the behavior of the symmetry energy coefficients (s.e.c., 
which can also be called generalized screening functions for each channel)  
from 0.5 $\rho_0$ up to 2 $\rho_0$.
The symmetry energy coefficient for some of the forces in Figure 1
reach a maximum value when
$\rho_0 < \rho \leq 1.7 \rho_0$ 
and then decrease. This last behavior is typical of non relativistic
calculations.
There are two forces which yield different
behaviors: SkSC4 makes the slope of $A_{0,1}$ much less negative
reaching an instability point ($A_{0,1} < 0$) 
at smaller densities than the other
forces and SkSC10 for which the slope is larger and positive. 
This last behavior is present in relativistic calculations.
The point in which the symmetry energy coefficient 
is zero would correspond
to a phase transition of nuclear matter to an asymmetric state 
of neutron or proton matter.
Although the usual Skyrme forces  are not well suited
for high densities there is a tendency of a decrease of this 
screening function (s.e.c.)
until a  phase transition in a dense nuclear system.
The dependence on $b$, on the other hand, leads to a more repulsive
behavior as seen in Figures 2 and 3 eventually compensating
the attractiveness of low or high densities.

In Figure 2 the same parameter, $A_{0,1}$, is shown as a function
of $\rho/ \rho_0$ for the 
interactions SLy, SkSC4 and SkSC6 
for non zero n-p asymmetries, i.e., for b=.25 and b=.54. 
The latter corresponds to a n-p asymmetry of the nucleus of $^{208}Pb$.
The absolute values of the isovector symmetry energy coefficient (s.e.c)
increase with relation to the symmetric nuclear matter but the
behavior with varying $\rho$ is nearly the same in general.
However, for the force SkSC6, 
the isovector screening function has a higher slope in this 
range of densities.
The same coefficient as a function of the
 n-p asymmetry b is shown in Figure 3
at the densities $0.5, 1$ and $2 \rho_0$. 
Again we note different 
behaviors depending on the force. 
For the forces
SLy and SkSC4 the s.e.c. increases, as noted above, 
 for higher densities and n-p asymmetries but for 
$\rho = 2 \rho_0$ the values are smaller at higher asymmetries.
For the force
SkSC6 the higher the density
the higher the slope with $b$. 
The tendency is that the interaction becomes more
repulsive.
Whereas the dependence of $A_{0,1}$ on $b$ is a new possibility 
its dependence on $\rho$ has been studied quite extensively and 
a distinction between the behavior of relativistic and non relativistic
models can be done which still are nearly observed here. 
While relativistic calculations lead to a continuous increase of 
the symmetry energy coefficient \cite{PAL}, non relativistic 
(microscopic or not) calculations usually yield a saturation at densities 
of the order of $2.5 \rho_0$ with a decreasing value at higher densities
\cite{PAL,BALDOBOMBURG}. 
Although these are the more common results there is a non relativistic
variational calculation  with three body forces \cite{BALDOBOMBURG}
which presents an increasing symmetry energy coefficient for higher
densities. 
All these calculations adopt the parabolic approximation for the
symmetry energy as discussed by \cite{HUBERWEBERWEIGEL} 
\footnote{We leave for a forthcoming work 
a discussion about corrections to the parabolic approximation
in the frame of a relativistic model,
following ideas contained in section 2 of the 
present work and in \cite{ISOSYMEN}.}.
We have just shown, however, that it is possible to obtain 
the behavior typical from relativistic calculations with
the Skyrme force SkSC10.

The spin symmetry energy coefficient is plotted in Figures 4, 5 and 6
as a function of $\rho/\rho_0$ and $b$ respectively.
In Figure 4 one sees that, for increasing densities, 
$A_{1,0}$ may have different slopes.
For the interactions SkSC4, SkSC6 and SkSC10 the spin- symmetry 
energy coefficient is already negative at low densities
and decrease still more with $\rho$.
This feature is clear in Figure 5 where the same
spin s.e.c. is shown for non zero n-p asymmetries.
These forces exhibit a behavior typical of a ferromagnetic
matter with stronger spin alignment as density increases.
With the force SLy, on the other hand, the spin symmetry 
coefficient indicates that no polarized state occurs the range
of densities under study.

In Figure 6 the spin symmetry energy is shown as a function
of $b$ with forces SLy, SkSC4 and SkSC6 for some different 
densities. The common trend is the increase of $A_{1,0}$
with $b$, i.e., at very asymmetric matter the spin interaction tends to 
become more repulsive. 
However the particular behavior of the spin s.e.c. with $b$ is different for
each effective force at a given density.
This means that whereas for the SLy force the increase of n-p asymmetry 
makes $A_{1,0}$ to increase (proportionally) at each density considered, 
for the forces SkSC4 and SkSC6 the slope is higher for higher densities.
This makes the curves of $A_{1,0}$ for different densities to cross at certain 
n-p asymmetry.
We can compare our results to the ratio of spin susceptibility  
of interacting neutron matter (for which $b \to \infty$) 
to the non interacting Fermi gas obtained by
Fantoni, Sarsa and Schmidt \cite{FSS2001} by means of the 
auxiliary field diffusion Monte Carlo method. 
This ratio is proportional to the polarizability as obtained in expression
(\ref{20}) and therefore inversely proportional to the spin
symmetry coefficient $A_{1,0}$. 
First of all we note that, in most cases, contrarily to what we find,
the values they find are 
all positive for the range of densities considered by them, from $0.75 \rho_0$ 
up to $2.5 \rho_0$. 
However the slope seems to be nearly the same as that we obtain
for low values of the n-p asymmetry. 
Consequently they may obtain
instabilities for higher density neutron matter whereas  we do not 
observe this result in our calculations with Skyrme forces 
(the comparison is meaningful for neutron
matter: $b$ is very large). Other comments about this are 
drawn in section 5.

The spin-isospin symmetry energy coefficient for
symmetric matter ($b=0$) as a function of $\rho/\rho_0$ 
is shown in Figure 7.
For all the forces,  $A_{1,1}$ 
are  decreasing 
functions of the density and they
become negative between  $\rho \simeq 1.8 \rho_0$ and $2.1 \rho_0$.
The only significant difference comes from the value of 
$A_{1,1}$ at low densities, being that SLy force 
implies a larger value than the SkSC forces.
In Figure 8 the same analysis holds for asymmetric matter for
$b=.25$ and $.54$. 
However, with increasing asymmetry, the instability
point is changed differently for each of the forces.
This is very interesting as well and very 
similar to the spin channel (of figure 5).
The SLy force presents the spin-isospin coefficient
which changes much less than the SkSC forces.
For these Skyrme interactions instabilities associated to pion condensation 
would be found already at low densities, $A_{1,1} < 0$ 
.
Therefore, for the  Skyrme type interactions we use,
we expect a there is a connection between the 
appearance of spin instabilities (leading to ferromagnetic states)
and spin-isospin instabilities (for pion condensation),
even  for not very asymmetric nuclear matter (up to $b \simeq 0.54$).

In Figure 9, $A_{1,1}$ is plotted as a function of $b$ for 
different Skyrme forces.
In this Figure the behavior of the spin-isospin 
s.e.c. with the n-p asymmetry parameter b is shown for 
different nuclear matter densities (below, equal to and above the 
saturation density). 
It can be noted that for the force SLy
an increasing n-p asymmetry (b) makes the spin-isovector
interaction continuously more repulsive instead of allow for
attractiveness while the opposite behavior is found for SkSC6. 
One would find the instabilities associated to pion condensation 
at roughly any nuclear density for SkSC6 at nearly any n-p asymmetry, 
which  does not seem to be realistic \cite{MIGDALrep}.
The other forces SkSC also reproduce this last behavior with different
slopes.
We are lead to conclude that  Skyrme interactions may be able to reproduce
roughly well  any kind of
dependence of the (generalized) symmetry 
energy coefficients in a particular range of density and/or n-p asymmetry.

In Figures 10 and 11 the scalar s.e.c. is shown as 
 a function of $\rho/\rho_0$ for b=0 and b=0.25 and .54 respectively.
These Figures show that, roughly speaking, 
all the interactions
are attractive or weakly repulsive at low densities and more repulsive
as density increase. 
This may be related to 
the instabilities found, for example, in \cite{AYCOCHO}.
However it is interesting to see in Figure 11 that,
depending on the interaction, 
with the increase of the n-p asymmetry  the scalar 
symmetry energy coefficient is reduced.
This occurs for lower densities but not necessarily at higher $\rho$. 
This is well seen in Figure 12 for $A_{0,0}$ as a function
of $b$. 
This may be understood as a general tendency
of resulting instabilities for (very) asymmetric nuclear matter
with a curious exception for force SLy at a density
$2 \rho_0$.
The behavior of the incompressibility with increasing n-p asymmetries,
at the saturation density, for the force SLy 
is nearly the same as that found in other calculations with Skyrme forces
or relativistic calculation \cite{BCK,YST} and 
microscopic variational study at finite temperatures \cite{variationalK}.
For a higher density ($2 \rho_0$), however, 
this scalar coefficient increases with $b$ 
for the  same SLy force.
A comparison of this kind was explicitly done in \cite{ISOSYMEN}.
The other forces exhibit smaller slopes and make 
$A_{0,0}$ to decrease faster. 
This last kind of behavior, more pronounced, was 
also found in other works (for the bulk incompressibility 
modulus $K_{\infty}$ depending on the 
used interaction \cite{YST,variationalK}.
Therefore the behavior of the incompressibilities ($K_{\infty}$ and 
$A_{0,0}$ - the "dipolar incompressibility") 
with n-p asymmetry depend strongly on the used interaction.

\section{ Stability of Asymmetric nuclear matter}

Consider that the binding 
energy can be minimized with respect to $\beta$ yielding an equilibrium 
state for nuclear matter. 
This means that: $d (E/A)/ d \beta = 0$, yielding
a differential equation for $\cA_{s,t}$ as a function of $\beta$ or $\Pi_{s,t}$
by expression (\ref{1bs}).
In the case $\rho$ is not dependent on $\beta$ 
the solution for this resulting differential equation 
is given by \cite{ISOSYMEN}:
\be \label{solucao} \ba{ll}
\displaystyle{ \frac{\cA_{s,t}}{ \rho_0} = \frac{C}{\Pi_{s,t}^2} 
- \frac{1}{\Pi_{s,t}}
,}
\ea
\ee
where $C$ is a constant.
Inverting the above expression we find:
\be \label{PI-A} \ba{ll}
\displaystyle{ \Pi_{s,t} = - \frac{ \rho_0}{2 \cA_{s,t}}
\pm \sqrt{  \frac{4 C_{s,t} \rho_0}{ \cA_{s,t} } + 
\frac{\rho_0^2}{4 \cA_{s,t}^2 } }, }
\ea
\ee
which, in principle, can be either real or complex.
Furthermore, there may appear two branches of solutions 
for $b \neq 0$. We want to emphasize that the relation
between this expression and those obtaining from
the linear response with Skyrme forces is not 
completely understood.
As a boundary condition one requires the usual
expression for the symmetric limit obtaining:
\be \label{Const} \ba{ll}
\displaystyle{ C^{(s,t)} = - \frac{\rho_0}{4\cA_{s,t}^{sym}}. }
\ea
\ee
Such that the corresponding polarizability 
of symmetric nuclear matter $\Pi^{s,t}_0$ by: 
$\Pi_0^{s,t} = - \rho_0/(2 \cA^{sym}_{s,t})$ \cite{BVA}.

To really be a stable minimum of the binding energy the 
second derivative of the energy must be positive. 
Using the form given by expression (\ref{1bs}), and the 
approximation that $\rho_0$ does not depend on $\beta$, we can write
this second derivative as:
\be \label{secder} \ba{ll}
\displaystyle{ \frac{d^2 H}{d \beta^2} = -\frac{1}{\Pi}
+ \frac{4 \Pi}{\rho_0} \frac{d \cA}{d \Pi} + \frac{\Pi^2}{\rho_0}
\frac{d^2 \cA}{d \Pi^2}. }
\ea
\ee
With the 
solution given by expression (\ref{solucao}) we find, in the $(s,t)$
channel, that:
\be \label{cond} \ba{ll}
\displaystyle{ \frac{d^2 H}{d \beta^2_{s,t}} = 
- \frac{\cA_{s,t}}{\rho} - \frac{C}{\Pi^2_{s,t}} =
\pm \frac{2 \sqrt{\rho^2 + 4C\rho_0 \cA_{s,t}}}{\rho \; \Pi^{s,t}} >0 
.}
\ea
\ee
The constant $C$ may be negative (stable symmetric nuclear matter 
according to expression (\ref{Const})) or positive and 
$\cA$ and $\Pi$ also may have negative or positive signs.
It is possible to have a case where the symmetric nuclear
matter is unstable, $\cA_{sym} < 0$, but a stable 
asymmetric nuclear matter is obtained, 
with positive $d^2 H/d \beta^2$.
This is a curious result. 
It seems to suggest that the stability line of nuclei
can occur, at least with its qualitative real behavior, 
without introducing explicitly electromagnetic forces. 

For the sake of comparison we quote the usual 
stability condition for a Fermi liquid
in each channel of the interaction. It  is  given by:
\be
a_{s,t} = N_0 ( 1+ J_0^{s,t}) > 0.
\ee
 where $J_0^{s,t}$ 
stands for any of $F_0, F'_0, G_0, G'_0$ respectively for 
the scalar ($s=0,t=0$), 
isovector ($s=0,t=1$), spin ($s=1,t=0$) and 
spin-isovector ($s=1,t=1$) channels \cite{FERLIQ}. 
These conditions correspond to the denominators of the response function
of {\it symmetric} nuclear matter at zero temperature and at saturation
density.

\section{Further discussion and Summary}

The expressions we have derived and used for the (isospin) symmetry
energy coefficient (with and without effective Skyrme forces) 
break isospin symmetry because the results for 
an excess of neutron (density) is different from those of 
higher proton fraction. 
In principle this is expected although 
usually considered to be small in mass formulae for finite nuclei 
\cite{MONI}. 
With the expressions we derived the effects of this symmetry 
breaking can be expected to 
be larger eventually related to other effects \cite{MIXING}.

Recently it has been proposed that in the core of dense (neutron) stars
there may occur a phase of matter in which color degrees of freedom are
deconfined generating di-quark condensates in channels in which the 
interaction is attractive. 
Several possibel scenarios for this color superconducting phase of matter
can be drawn, among them the color-flavor locked phase \cite{CFL}.
We would like to point out that with the increase of the 
isospin symmetry energy coefficient, for instance, the resulting 
fraction of protons, after the supernovae stage, would be higher 
hindering the emergency of a pure neutron matter in the death of the star.
As a consequence the final proportion of up and down
quarks would be different than that from a pure neutron star.
This may modify the color superconducting phase.
These remarks may apply to relativistic or high energy
heavy ions collisions which generates high baryonic density.

The relevance of the isovector symmetry energy coefficient and its 
dependence on the nuclear density has been intensively studied due to its
relevance to the determination of the nuclear matter equation of state 
which is very important for the comprehension of high energy 
ions collisions and eventually the dynamics of astrophysical objects.
An interesting recent study was done by Bao An Lin \cite{BAO} who 
however only is concerned about the parabolic approximation without 
a more general form of the asymmetry term. This extension would be 
of high interest for the field.

Summarizing, the dependence of the s.e.c. on the n-p asymmetry
was studied extending the results of ref. \cite{ISOSYMEN,BJP2003}.
Generalized symmetry energy coefficients
of asymmetric (non relativistic) nuclear matter at 
variable densities were investigated.
The density and 
n-p asymmetry dependences of the s.e.c. in the different channels
were analyzed for
different Skyrme forces. 
They may yield very different
behaviors including the possibility  (or not) 
of nuclear matter to undergo phase transitions.
These forces can describe different behaviors
of the symmetry energy coefficients.
Although one should not believe that only one  
Skyrme force parametrization could account for the description
of all nuclear observables at different densities, temperatures and 
n-p asymmetries it is acceptable the idea that several parametrizations
could hopefully describe different ranges of the dependence of 
nuclear observables (as the s.e.c.)
with these variables.
Whereas the isovector, scalar and spin channels were rather an extension
of the work presented in \cite{ISOSYMEN,BJP2003} the results for 
spin and spin-isospin channels are scarce in the literature.
We showed that, for the Skyrme forces under consideration (SLy,
SkSC4, 6, 10) there is a kind of correlation between the two channels
and the possibility of occurring  simultaneous 
ferromagnetic instabilities and/or 
pion condensation at high densities and n-p asymmetries.
Therefore, in principle, different values can be expected  for the 
(bulk) symmetry energy coefficients in nuclei and nuclear matter, 
with different n-p asymmetries, and neutron stars. 
In the spin channels  it is possible to expect
spin polarized asymmetric matter 
yielding magnetic fields in neutron stars, as discussed
in \cite{KUTSCHERA}, according
to the results in figures 4,5 and 6 for forces the SkSC.
However with the increase of $b$ we find that
the spin interaction may be rather repulsive, hindering
this magnetization effect with the use of these Skyrme forces
in the range of densities analyzed here.
A similar behavior for the spin-isospin channel 
(which may indicate instabilities for pion condensation) 
was found with the same Skyrme interactions.
A  new stability condition
for an asymmetric nuclear medium was proposed.
It is given by:
$$ \frac{d^2 H}{d \beta^2_{s,t}} = 
- \frac{\cA_{s,t}}{\rho} - \frac{C}{\Pi^2_{s,t}} =
\pm \frac{2 \sqrt{\rho^2 + 4C\rho_0 \cA_{s,t}}}{\rho \; \Pi^{s,t}} >0 .$$
Where
 $C$ is a constant  and $\Pi^{s,t}=\rho/(2 \cA_{s,t})$ for each channel.

\vskip 0.3cm
\noindent {\Large {\bf Acknowledgements}}

This work was supported by FAPESP, Brazil.

\vspace{2cm}

\noindent {\Large {\bf  Figure captions}}

\vskip 0.5cm

\noindent {\bf Figure 1 } Neutron-proton symmetry energy coefficient 
 $A_{0,1}= \rho/(2 \Pi_R^{0,1})$ of symmetric nuclear matter
 as a function of the ratio
of density to the saturation density ($u=\rho/\rho_0$) for interactions SLyb
(dotted-dashed line),
SkSC4 (solid), SkSC6 (dotted), SkSC10 (dashed).

\noindent {\bf Figure 2} The same as Figure 1 
for interactions SLy
with b=0.25 (thick dotted-dashed line), 
b=.54 (thin dotted-dashed line), SkSC6 with b=.25 (thick dotted),
b=.54 (thin dotted), SkSC4 with b=.25 (thick solid) and b=.54 (thin solid).

\noindent {\bf Figure 3}  Neutron-proton symmetry energy coefficient 
$A_{0,1}= \rho/(2 \Pi_R^{0,1})$ as a function of the
asymmetry coefficient $b$ at different densities: 
solid lines for SkSC4 (very thick: $\rho=.5\rho_0$,
thick:$\rho=\rho_0$, thin:$\rho=2\rho_0$,), dotted lines for SkSC6 
(same conventions as SkSC4) and dotted-dashed lines for SLy.

\noindent {\bf Figure 4}  Spin symmetry energy coefficient 
$A_{1,0}= \rho/(2 \Pi_R^{1,0})$ as a function of the ratio
of density to density at saturation ($u=\rho/\rho_0$) for interactions 
SLy, SkSC4, SkSC6 and SkSC10 with the same conventions of Figure 1.

\noindent {\bf Figure 5}  Spin symmetry energy coefficient 
$A_{1,0}= \rho/(2 \Pi_R^{1,0})$ as a function of the ratio
of density to density at saturation ($u=\rho/\rho_0$) for interactions 
SLy, SkSC4, SkSC6 and SkSC10 with the same conventions of Figure 2.

\noindent {\bf Figure 6}  Spin symmetry energy coefficient 
$A_{1,0}= \rho/(2 \Pi_R^{1,0})$ as a function of the asymmetry
coefficient b at different densities and for the different forces:
with the same conventions of Figure 3.

\noindent {\bf Figure 7}  Spin-isospin symmetry energy coefficient 
$A_{1,1}= \rho/(2 \Pi_R^{1,1})$
as a function of the ratio
of density to density at saturation ($u=\rho/\rho_0$) for interactions 
SLy, SkSC4, SkSC6 and SkSC10 with the same conventions of Figure 1.

\noindent {\bf Figure 8} Spin-isospin symmetry energy coefficient 
$A_{1,1}= \rho/(2 \Pi_R^{1,1})$ as a function of the
ratio
of density to density at saturation ($u=\rho/\rho_0$) for 
the different interactions:
 with the same conventions of Figure 2.

\noindent {\bf Figure 9} Spin-isospin symmetry energy coefficient 
$A_{1,1}= \rho/(2 \Pi_R^{1,1})$ as a function of the asymmetry
coefficient b at different densities and with the different forces:
using the same conventions of Figure 3.

\noindent {\bf Figure 10} Scalar symmetry energy coefficient 
 $A_{0,0}= \rho/(2 \Pi_R^{0,0})$
as a function of the ratio
of density to density at saturation ($u=\rho/\rho_0$) for the
different interactions with the conventions of Figure 1.

\noindent {\bf Figure 11} Scalar symmetry energy coefficient 
$A_{0,0}= \rho/(2 \Pi_R^{0,0})$ as a function of the ratio
of density to density at saturation ($u=\rho/\rho_0$) for the
different  interactions:
with the conventions of Figure 2.

\noindent {\bf Figure 12}  Scalar symmetry energy coefficient 
 $A_{0,0}= \rho/(2 \Pi_R^{0,0})$ as a function of the asymmetry
coefficient $b$ at different densities and with the different forces,
the conventions of Figure 3.

\newpage

\begin{figure}[htb]
\epsfig{figure=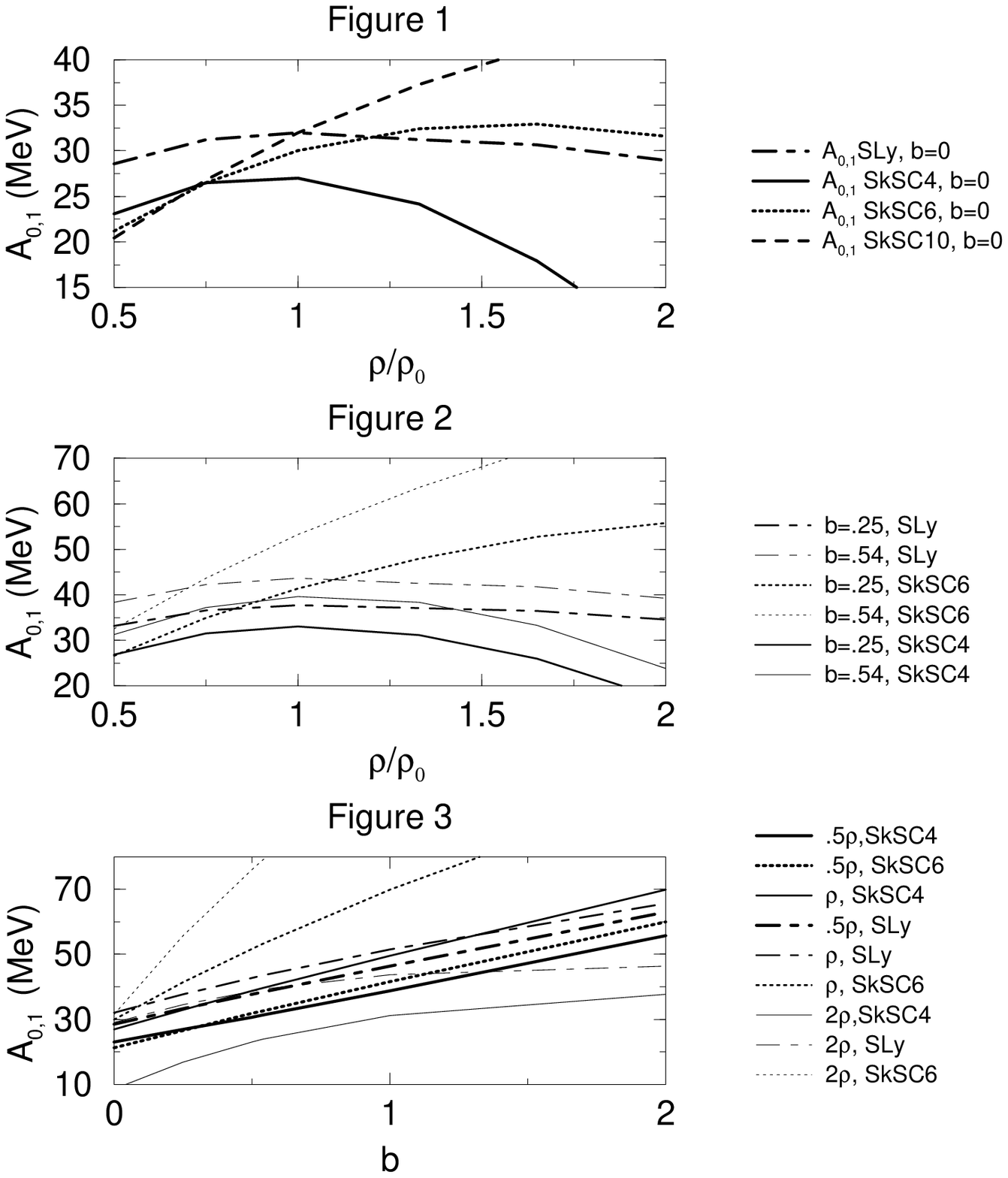,width=14cm} 
\end{figure}

\newpage

\begin{figure}[htb]
\epsfig{figure=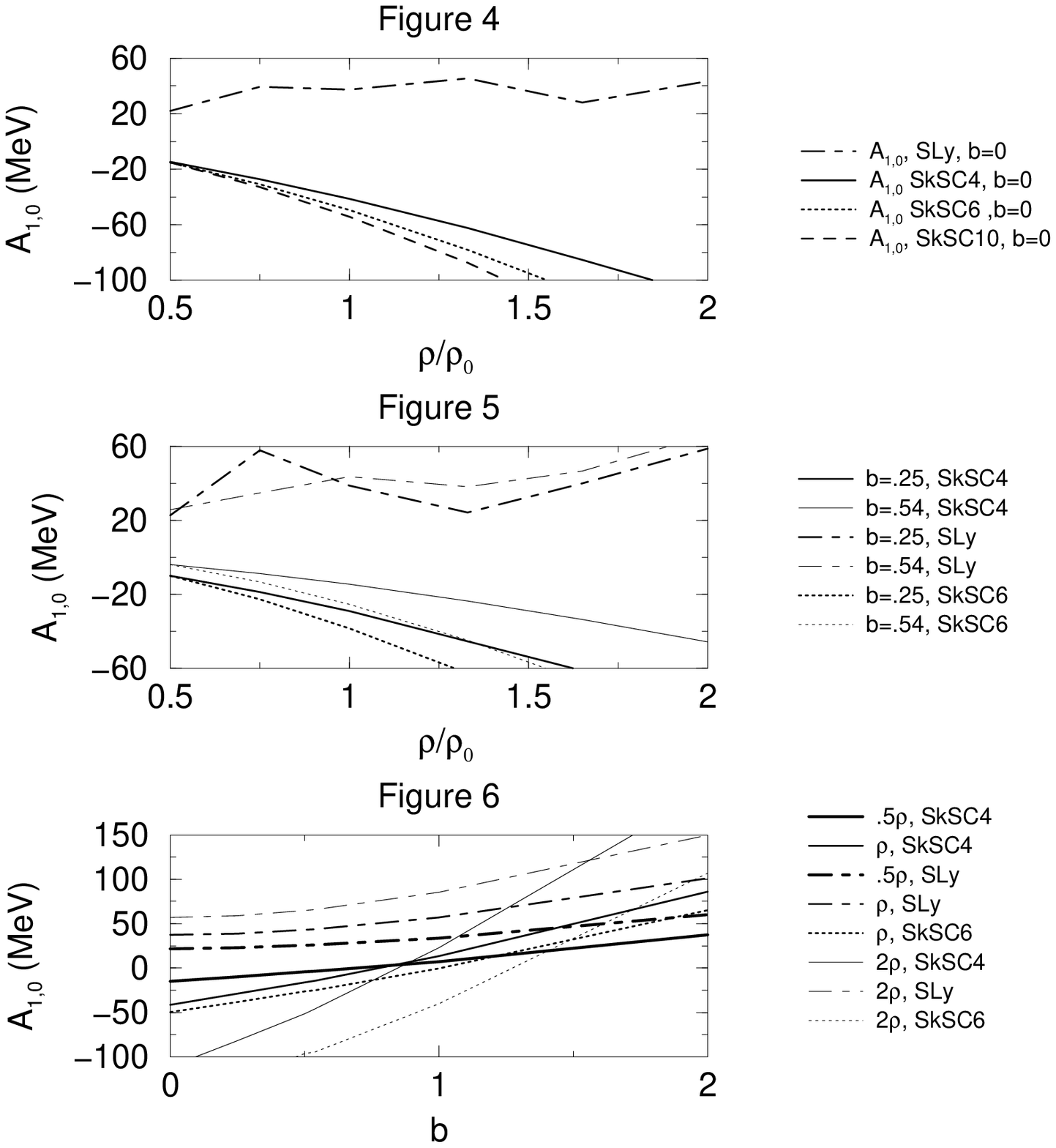,width=14cm} 
\end{figure}

\newpage

\begin{figure}[htb]
\epsfig{figure=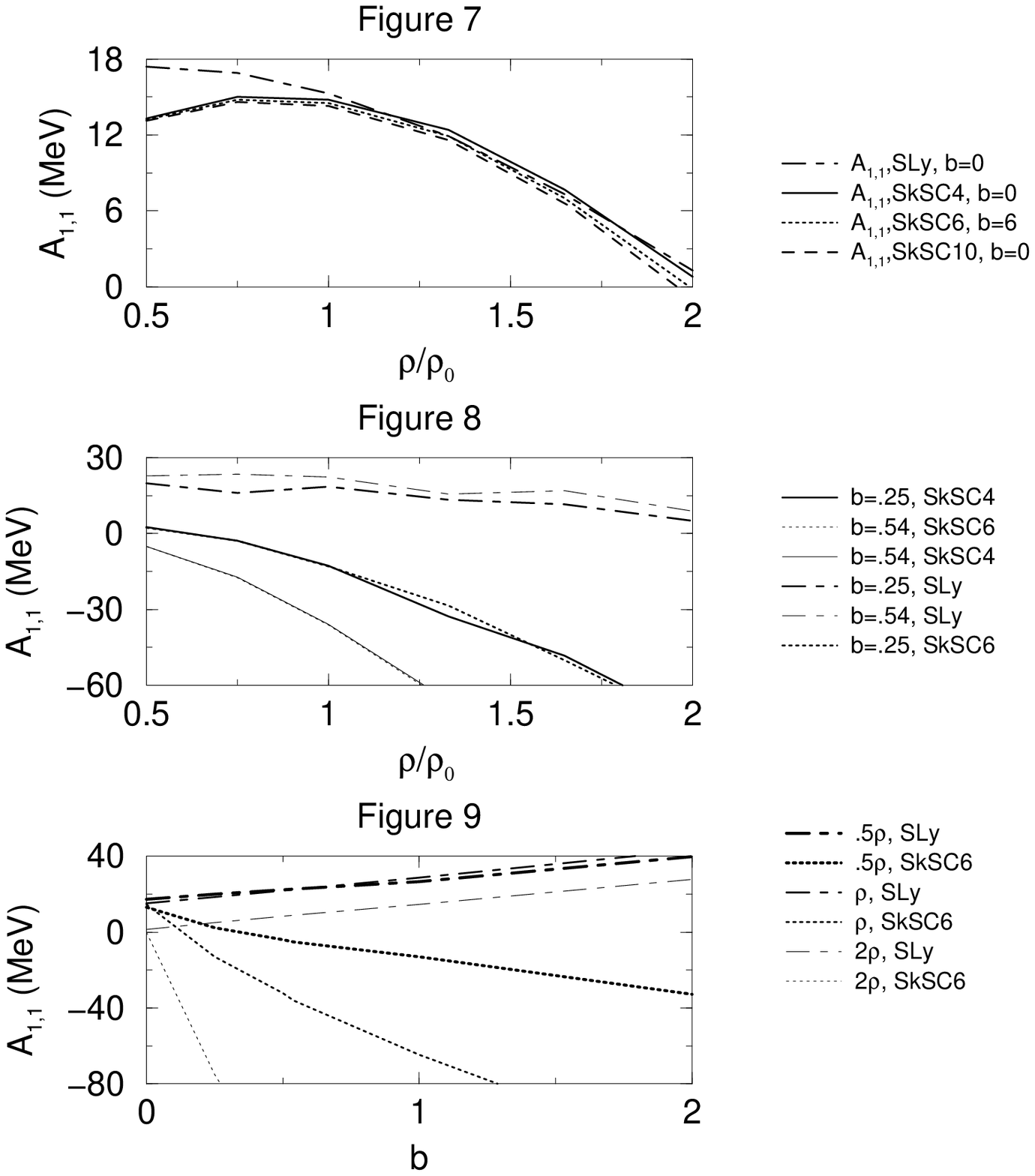,width=14cm} 
\end{figure}

\newpage

\begin{figure}[htb]
\epsfig{figure=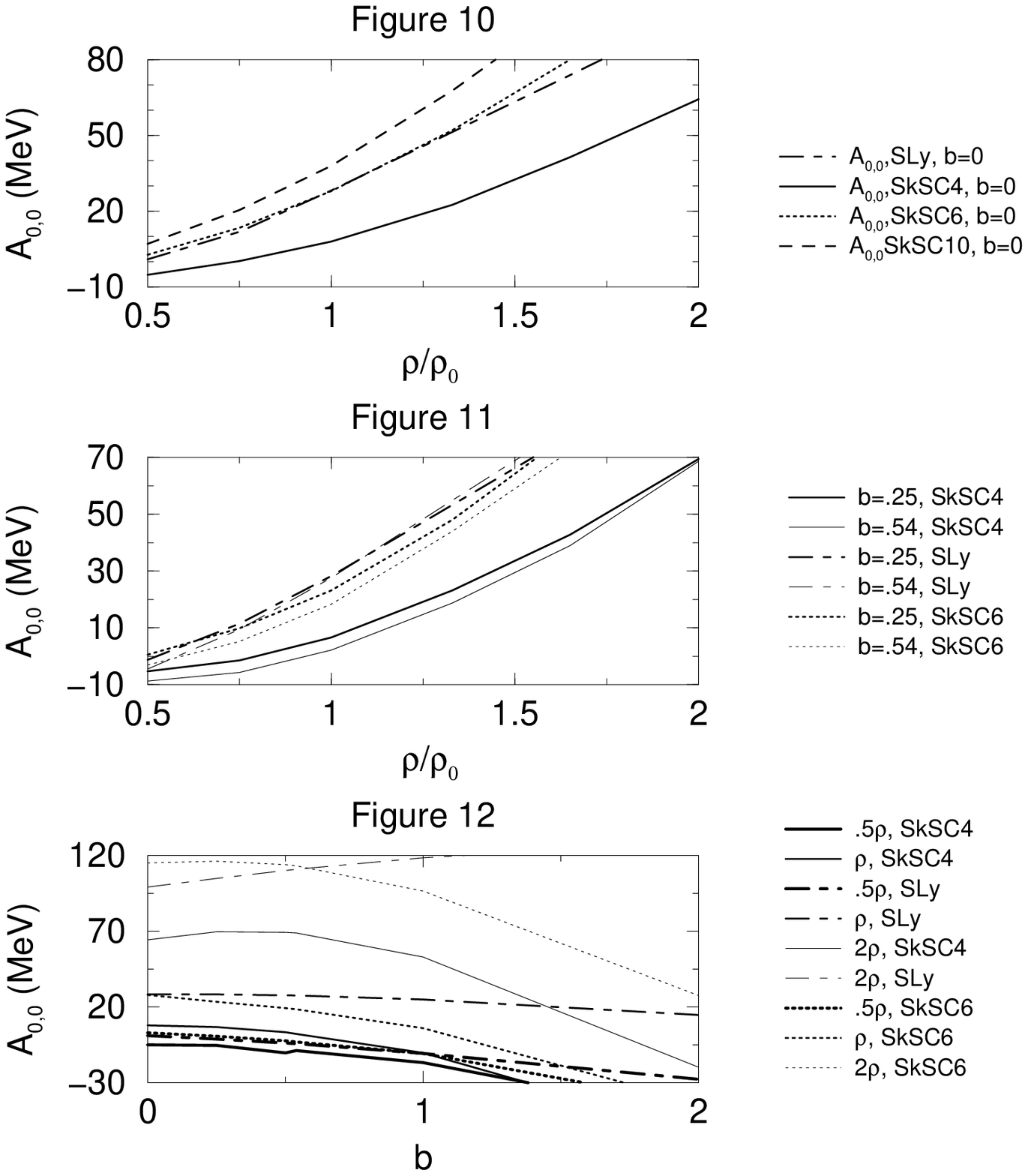,width=14cm} 
\end{figure}


\vskip 0.2cm



\begin{thebibliography}{ll}
\bibitem{JACO} J. J\"anecke and E. Comay, Nucl. Phys. {\bf A 436} 
(1985) 108.
\bibitem{LKLB} C.-H. Lee {\it et al}, Phys. Rev. {\bf C 57} (1998) 3488.
\bibitem{MONI} 
P. Moeller, J.R. Nix, W.D. Myers and W.J. Swiatecki,
At. Data Nucl. Data Tables {\bf 59}, 185 (1995);
Y. Aboussir, J.M. Pearson, A.K. Dutta and F. Tondeur,
At.  Data Nucl. Data Tables {\bf 61}, 127 (1995).

\bibitem{ISOSYMEN} F.L. Braghin, 
Nuc. Phys. {\bf A 696} (2001) 413; and  F.L.B., 
{\it Erratum} Nuc. Phys. {\bf A} to be published.

\bibitem{FLB99} F.L. Braghin, Phys. Lett. {\bf B 446}, (1999) 1;
 {\it idem ibid:  Erratum} 
submitted to P.L. B.; Nucl. Phys. {\bf A 665}, (2000) 13

\bibitem{BJP2003} F.L. Braghin, Brazilian Journal of Physics, accepted to 
publication, {\it Proceedings of the Brazilian Meeting of Nuclear Physics},
September 2002, S\~ao Paulo, ed. D.P. Menezes.

\bibitem{SAWYER} R.F. Sawyer, Phys. Rev. {\bf C 11}, 2740 (1975), 
N. Iwamoto and C.J. Pethick, Phys. Rev. {\bf D 25}, 313 (1982).
\bibitem{ESPANHOIS} J. Navarro, E.S. Hernandez, D. Vautherin, 
Phys. Rev. {\bf C 60} (04)5801 (1999).
\bibitem{REDDY} S. Reddy, M. Prakash, J.M. Lattimer, J.A. Pons,
  Phys. Rev. {\bf C 59}, (1999) 2888.

\bibitem{KUTSCHERA} M. Kutschera and W. W\'ojcik, Phys. Lett. {\bf B 223},
11 (1989)


\bibitem{CHABANAT} E. Chabanat {\it et al},  Nucl. Phys. {\bf A 627}, 
 (1997) 710.
\bibitem{DUTOABO} A.D. Dutta {\it et al}, Nucl. Phys. {\bf A 458},
77 (1986); F. Tondeur {\it et al}, Nucl. Phys. {\bf A 470}, 93 (1987),
Y. Aboussir {\it et al}, Nucl. Phys. {\bf A 549}, 155 (1992).
\bibitem{ONSIPP} M. Onsi, H. Przysiezniak and J.M. Pearson,
Phys. Rev. {\bf C 50}, (1994) 460.

\bibitem{BVA} F.L. Braghin, D. Vautherin and A. Abada, Phys. Rev. {\bf C
52},  (1995) 2504.

\bibitem{PAL} M. Prakash, T.L. Ainsworth, J.M. Lattimer, Phys. Rev. Lett.
  {\bf 61} 2518 (1988) 


\bibitem{BALDOBOMBURG} M. Baldo, I. Bombaci, G.F. Burgio, Astron. Astrophy.
 {\bf 328} 274 (1997).
\bibitem{HUBERWEBERWEIGEL} H. Huber, F. Weber, M.K. Weigel, Phys. Rev. {\bf C 51},
  1790 (1995).

\bibitem{FSS2001} S. Fantoni, A. Sarsa, K.E. Schmidt, Phys. Rev. Lett. {\bf 87},
181101 (2001)

\bibitem{MIGDALrep} A.B. Migdal, E.E. Saperstein, M.A. Troitsky, D.N. Voskresensky,
      Phys. Rep. {\bf 192} 179 (1990).

\bibitem{AYCOCHO} S. Ayik, M. Colonna, Ph. Chomaz, Phys. Lett.
{\bf B 353}, (1995) 417.

\bibitem{BCK} E. Baron, J. Cooperstein, S. Kahana, Phys. Rev. Lett. {\bf 79}
(1997) 609.
\bibitem{YST} S. Yoshida. H. Sagawa and N. Takigawa, Phys. Rev. {\bf C 58}, 2796
(1998)

\bibitem{variationalK} M. Modarres and G.H. Bordbar, Phys. Rev. {\bf C 58}, 2781
(1998)


\bibitem{FERLIQ} D. Pines and P. Nozieres, {\it The theory of Quantum 
Liquids}, W.A. Benjamin Inc., New York, 1966.

\bibitem{MIXING} For example: L.A. Barreiro, A.P. Gale\~ao, G. Krein, 
   Phys. Lett. {\bf B 358}  7 (1995); W. Broniowski, W. Florkowski, 
  Phys. Lett. {\bf B 440}  7 (1998).

\bibitem{CFL} For example: M. Alford, J.A. Bowers, F. Rajagopal,
       Phys. Rev. {\bf D 63} 074016 (2001).

\bibitem{BAO} Bao- An Lin, Phys. Rev. Lett. {\bf 88}, (2002) 192701; 
Nucl. Phys. {\bf A 708}, (2002) 365.


\end{thebibliography}
\end{document}